\documentclass[%
 reprint,
superscriptaddress,
preprintnumbers,
 amsmath,amssymb,
 aps,
]{revtex4-1}
\pdfoutput=1
\usepackage{graphicx}
\usepackage{dcolumn}

\newcommand{\GeV}{\text{\,GeV}}

\newcommand{\be}{\begin{equation}}
\newcommand{\ee}{\end{equation}}
\newcommand{\ba}{\begin{eqnarray}}
\newcommand{\ea}{\end{eqnarray}}

\renewcommand{\l}{\left(}
\renewcommand{\r}{\right)}

\newcommand{\Br}{\mathop\mathrm{Br}}

\begin{document}

\preprint{INR-TH/2014-026}

\title{Decaying light particles in the SHiP experiment: \\ 
Signal rate estimates for hidden photons}

\author{D. Gorbunov}
 \email{gorby@ms2.inr.ac.ru}
\affiliation{Institute for Nuclear Research of the Russian Academy of Sciences,
  Moscow 117312, Russia}%
\affiliation{Moscow Institute of Physics and Technology, 
  Dolgoprudny 141700, Russia}%
\author{A. Makarov}
 \email{a.mr@mail.ru}
\affiliation{Institute for Nuclear Research of the Russian Academy of Sciences,
  Moscow 117312, Russia}%
\author{I. Timiryasov}
 \email{timiryasov@inr.ac.ru}
\affiliation{Institute for Nuclear Research of the Russian Academy of Sciences,
  Moscow 117312, Russia}%
\affiliation{Physics Department, Moscow State University, Vorobievy Gory,  
Moscow 119991, Russia}

\begin{abstract}
For the extension of the Standard Model with light hidden photons 
we present preliminary estimates of the signal rate expected at the
recently proposed fixed target SHiP experiment 
exploiting the CERN SPS beam of 400 GeV protons. 
\end{abstract}

\maketitle

\section{Introduction: the experiment and the model to be tested
\label{sec:intro}}

Unsolved phenomenological problems---neutrino oscillations, dark
matter phenomena, baryon asymmetry of the Universe---definitely ask
for an extension of the Standard Model of particle physics (SM). It is
natural to find the corresponding new particles at a mass scale not
much higher than the electroweak scale. Otherwise the hierarchy
problem arises in the scalar sector: quantum corrections of heavy
particles push the SM Higgs boson mass up to their mass scale. While
the LHC scrutinizes thoroughly the (sub-)TeV scale, there is a logical
possibility of having thus far elusive new physics at a (much) lower
scale. The absence of any direct evidence of the new physics may be
attributed to the weakness of interaction between known and new
particles. In the search for such new physics the superior experiments are  
those operating on the high-intensity frontier.  

An example of this type of experiment is SHiP (Search for Hidden
Particles \cite{SHiP}), the recently proposed \cite{Bonivento:2013jag}
new fixed target experiment at the CERN SPS 400 GeV proton beam. The
original motivation \cite{Gninenko:2013tk} was to search for ${\cal
  O}(1)$\,GeV sterile neutrinos of $\nu$MSM, one of the most economic
extensions of the SM capable of explaining all the three
aforementioned phenomenological problems with only three new fields
(singlet with respect to SM gauge groups fermions) added to the SM, 
see \cite{Boyarsky:2009ix} for
review.  Mixing between singlet fermions and active neutrinos is
responsible for both the singlet production in decays of heavy mesons
(generated by protons on target) and subsequent singlet decays into SM
particles (the main signature for the SHiP detector), see
\cite{Gorbunov:2007ak} for details. The flux of secondary particles
from proton scatterings is suppressed by the very dense (tungsten)
dump placed downstream.  The main idea is to have a large detector  
($5\times$5\,m$^2\times$50\,m \cite{Bonivento:2013jag}) and place it
as close to
the target as possible (at a distance of about 60\,m 
\cite{Bonivento:2013jag}) in order to maximize the number of potential
singlet decays within the detector fiducial volume and still have the
background under control. This makes SHiP {\em a universal tool to probe
any new physics which introduces sufficiently light and long-lived
particles produced by protons on target and then decaying into SM
particles.} 

\begin{figure}[!htb]
\hskip 0.05\columnwidth
 \includegraphics[width=0.9\columnwidth]{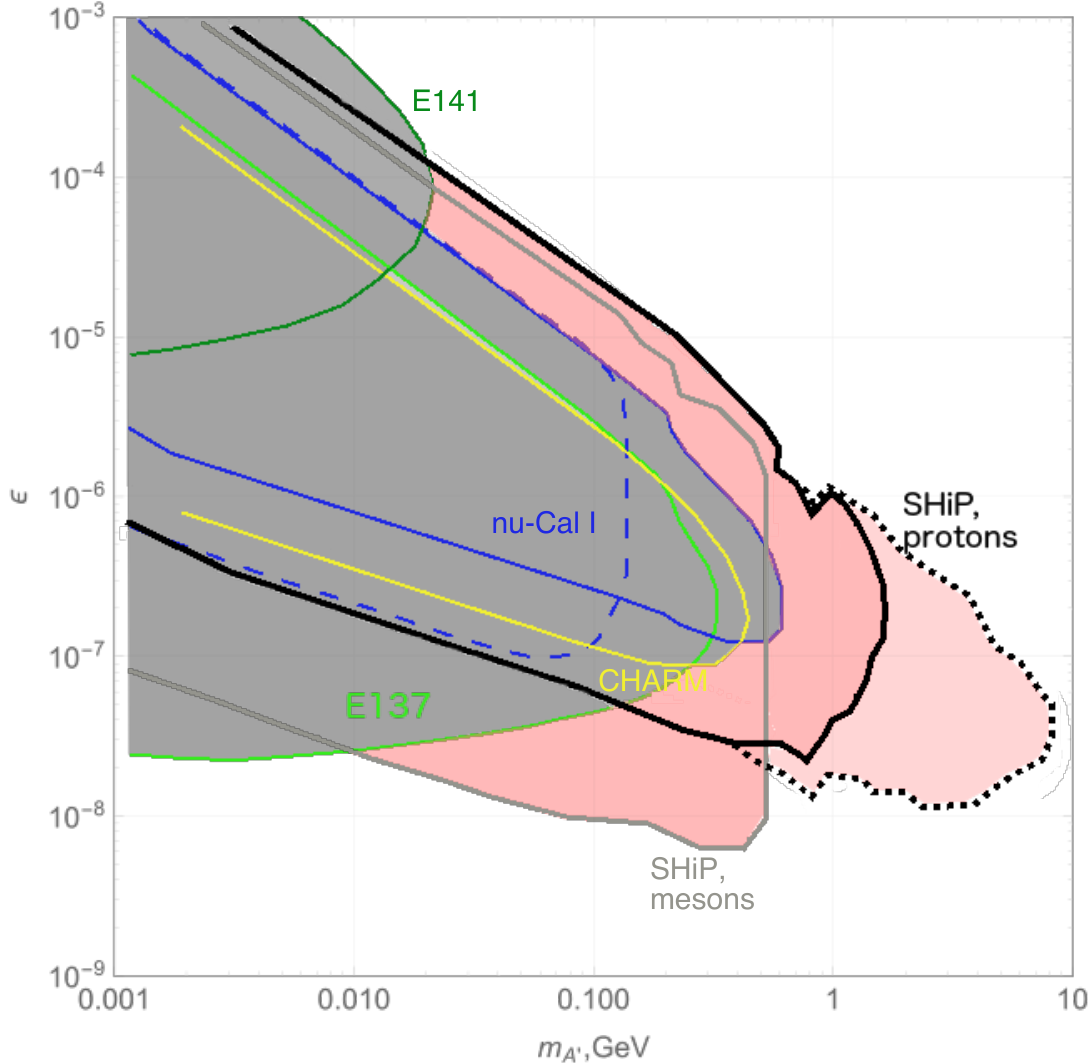}
\caption{Excluded regions in the hidden photon model parameter space   
(see \cite{Gninenko:2012eq,Andreas:2012mt,Blumlein:2013cua} for 
details), the   
 pink regions are expected from SHiP. Two estimates of proton
 contribution, with (solid line) and without (dotted 
 line) accounting for the proton form factor, are 
presented.  The correct line 
 goes in between, see Sec.\,\ref{subsec:proton}. 
\label{limits}}
\end{figure}

In this paper we consider one of the examples of such new physics,
which provides long-lived light particles that can be searched for at
SHiP, specifically models with massive hidden photons. 
The SM Lagrangian $\mathcal{L}_{SM}$ is extended in the following way:
\begin{equation}
\mathcal{L}=\mathcal{L}_{SM}-\frac14 F'_{\mu\nu}F'^{\mu\nu}+
\frac{\epsilon}{2}F'_{\mu\nu}F^{\mu\nu}+
\frac{m_{A'}^2}{2}A'_\mu A'^\mu,
\end{equation}
where $A'_\mu$ is massive gauge field of a new 
(dubbed dark) $U'(1)$ group, 
$F'_{\mu\nu}\equiv\partial_\mu A'_\mu-\partial_\nu A'_\mu$, 
and $\epsilon$ is parameter of kinetic mixing. 
The mixing provides effective coupling between $A'$
and pairs of the SM charged particles, which determines the model's 
phenomenology. 
Present phenomenological limits on $\epsilon$ and $m_{A'}$ are shown
in Fig.\,\ref{limits}. 
Hidden photon may be a messenger of the hidden sector,
that is responsible for (some of) the unsolved problems we started
with (see e.g. \cite{Pospelov:2007mp} for dark matter). 
The purpose of this work is to estimate the number of hidden photon
decays inside the SHiP detector. This is the first step
towards the estimate of the SHiP sensitivity to models with 
hidden photons, which in order to be completed,   
requires fixing the experiment 
layout, understanding the detection efficiency, and calculating the 
expected background.

\section{Production mechanisms\label{sec:hidden photon-production}}

Hidden photons can be produced directly, via proton (quark) or lepton
bremsstrahlung, and indirectly, in meson decays. 
The relevant leptons and mesons
are secondary particles produced either at proton scattering off
target or when the hadron and lepton cascades, initiated by
scattering, propagate in the dump material.

\subsection{Proton bremsstrahlung}
\label{subsec:proton}
In a fixed target experiment, particles $A'$ are generated by
scattering protons through a process analogous to ordinary photon
bremsstrahlung. Consider a proton of mass $m_p$ with initial 3-momentum
$P$ and initial energy $E_p$. Let $E_{A'}$ be the energy of $A'$ and 
$z$ denote a fraction of the momentum
$P$ carried away by $A'$ in the direction of incoming
proton. Therefore 
$P\cdot z = p_\parallel$, $p_\parallel$ and $\overrightarrow{p}_\perp$
are longitudinal and transverse components of $A'$ 3-momentum
$P_{A'}$. The 
differential $A'$-production rate per proton interaction, calculated in the
Weizs\"{a}cker-Williams approximation, reads \cite{Blumlein:2013cua} 
\begin{equation}
\frac{d N}{dz dp_\perp^2}=\frac{\sigma_{pA}(s')}{\sigma_{pA}(s)}\,
w_{ba}(z,p_\perp^2)\,,
\label{prod.rate}
\end{equation}
where $s'=2m_p(E_p-E_{A'})$, $s=2m_pE_p$ and  
\begin{multline*}
w_{ba}(z,p_\perp^2) = \frac{\epsilon^2 \alpha_{QED}}{2\pi H}
\left[\frac{1+(1-z)^2}{z} \right. \\ \left. 
-2z(1-z)\left(\frac{2m_p^2+m_{A'}^2}{H}
-z^2\frac{2m_p^4}{H^2}\right) \right. \\
\left.+2z(1-z)(1+(1-z)^2)\frac{m_p^2m_{A'}^2}{H^2}
+2z(1-z)^2\frac{m_{A'}^4}{H^2} \right]\,
\end{multline*}
with $H(p^2_\perp,z)=p_\perp^2+(1-z)m_{A'}^2+z^2m_p^2$ and fine
structure constant $\alpha_{QED}\approx 1/137$. The hadronic
cross section is factorized and related to the proton-proton
scattering cross section $\sigma_{pp}$ as 
$\sigma_{pA}(s) = f(A) \sigma_{pp}(s)$ with
function $f(A)$ depending only on atomic number $A$. Thus, it 
 drops out in expression \eqref{prod.rate} for the event rate. 
For inelastic proton-proton 
cross section we use the fit from \cite{Agashe:2014kda}.

Equation \eqref{prod.rate} was originally derived \cite{Kim:1973he}
under a set of specific conditions. For a beam-dump-type experiment 
these conditions could be summarized as follows
\cite{Blumlein:2013cua}:
\begin{equation}
E_p,\, E_{A'},\, E_p - E_{A'} \gg m_p, \, m_{A'}, \, \sqrt{p_\perp^2}.
\label{conditions}
\end{equation}
Another restriction comes from our treatment of scattering proton as
an entire particle and not a bunch of partons: we consider the proton, but not
quark, bremsstrahlung at high energies.  To ensure that we are dealing with the
entire proton, we restrict proton-nuclei momentum transfer, 
$(P-P_f-P_{A'})^2<\Lambda^2_{QCD}$, where $P_f$ denotes the 3-momentum
of the outgoing proton. Then, we also require the 3-momentum of the produced
$A'$ to be inside a cone determined by a detector geometry (see
Sec.\,\ref{sec:hidden photon-signal}). We have checked
numerically that the latter two restrictions guarantee the fulfillment
of conditions \eqref{conditions}. 
These two restrictions are
summarized in the following function:
\begin{equation}
f(z,p_\perp^2)=\theta\l\Lambda^2_{QCD}-(P-P_f-P_{A'})^2\r \cdot f_{\text{fiducial}}.
\label{fcut}
\end{equation}
with $f_{\text{fiducial}}$ referring to the restriction from the
detector geometry. Finally,  
to obtain a number of $A'$ whose trajectories cross the fiducial volume of
detector one integrates eq.\,\eqref{prod.rate} with factor 
\eqref{fcut}. 

In the considerations above we neglected the proton internal structure, 
however, the momentum transfer for heavy $A'$ emission can be
sufficient to feel it.  
This effect can be addressed at the parton level, which we leave for
future studies. Here, to be conservative, we restrict the kinematics
to the region where details of the internal structure are not
important, which is done by introducing the Dirac  $F_1$ and Pauli
$F_2$ 
form factors into the proton electric current  
(see, e.g. \cite{Arrington:2006zm}):
\begin{equation*}
J_\mu(p,p')=\bar{u}(p') \l \gamma_\mu F_1(q^2)+\frac{1}{2m_p} i 
\sigma_{\mu\nu} q^\nu F_2(q^2) \r u(p),
\end{equation*}
where $q=p-p'$ is the 4-momentum transfer. 
Then the production rate of $A'$ \eqref{prod.rate} gets multiplied by 
$F^2_1(m_{A'}^2)$ (the contribution of $F_2$ can be neglected
since $F_1(q^2)/F_2(q^2) \sim q^2$ for large $q^2$).
In the simplest dipole parametrization the Dirac form factor has form 
$F_1=(1+q^2 / m_D^2)^{-2}$ with $m^2_D=12/r_D^2$ being the Dirac mass
squared and the Dirac radius $r_D\approx 0.8$\,fm 
\cite{Gockeler:2006uu}.  
The dependence on $q^2$ refers to a recoil of other constituents that is
 redundant since, in fact, all final hadronic states contribute. 
Hence, introduction of
the  proton form factor leads to sufficient underestimate of the $A'$
 flux compared to the quark bremsstrahlung.  {\it Given our
 restrictions we arrive at the conservative lower limit on the hidden
 photon flux.}

\subsection{Secondary particles bremsstrahlung}
Hidden photons can be created by secondary particles in the beam dump
(tungsten, lead, etc.). The 
$A'$-production cross section in the electron bremsstrahlung process
was calculated in \cite{Bjorken:2009mm} in the Weizs\"{a}cker-Williams
approximation. For an incoming electron of energy $E_0$, the
differential cross section to produce $A'$ of energy $E_{A'}\equiv x
E_0$ is
\begin{multline}
\frac{d\sigma}{dx \, d cos \theta_{A'}}\approx \frac{8
  Z^2\alpha_{QED}^3\epsilon^2E_0^2 \,
  x}{U^2}\frac{\chi}{Z^2}\\\times\left[(1-x+x^2/2)-\frac{x\,(1-x)m_{A'}^2
    E_0^2\,x\,\theta_{A'}^2}{U^2} \right],
    \label{ecrosssection}
\end{multline}
where $\theta_{A'}$ is the angle in the lab frame between the emitted
$A'$ and the incoming electron, Z is the atomic number of the dump 
atoms,
\begin{equation}
U=U(x,\theta_{A'})=E^2_0\,x\,\theta_{A'}^2+m_{A'}^2\,\frac{1-x}{x}+m_e^2\,x,
\end{equation}
and an effective flux of photons (emitted by a rapidly moving atom in
the rest frame of incoming electron) is defined as follows:
\begin{equation}
\chi\equiv\int_{t_{min}}^{t_{max}}\!dt\,\frac{t-t_{min}}{t^2}\,G_2(t),
\label{tinegration}
\end{equation}
where $t_{min}=(m^2_{A'}/2E_0)^2$, $t_{max}=m_{A'}^2$ and
$G_2(t)=G_{2,el}(t)+G_{2,in}(t)$ is a general electric form factor,
the sum of elestic and inelastic parts (see \cite{Bjorken:2009mm} for
details).  
To simplify numerical integration we neglect $x$- and
$\theta_{A'}$-dependences of  $t_{min}$ in \eqref{tinegration}. This
can lead  \cite{Beranek:2013nqa} to an overestimate of the cross section
by $\sim 30 \%$, which  only 
  insignificantly affects our results for the SHiP discovery potential.

\subsection{Production in meson decays}

Hidden photon $A'$ can emerge in meson electromagnetic 
decays (if kinematics admits) due
to mixing with photon.  
For the corresponding branching ratio of 
$\pi^0$ meson decay, one has the following estimate 
\cite{Batell:2009di}: 
\begin{equation}
\Br(\pi^0\to A' \gamma)\simeq 2\epsilon^2\left ( 1-
\frac{m_{A'}^2}{m_{\pi^0}^2} \right )^3 \Br(\pi^0\to\gamma\gamma)\,.
\label{neutral}
\end{equation}
There is a similar expression for $\eta^0$-meson with obvious
replacement $\pi^0\leftrightarrow\eta^0$ in \eqref{neutral}.

For branching ratios of vector meson $V$ (e.g., $V = \rho^\pm, \rho^0,\omega$)
decays into $A'$ and pseudoscalar meson $P$ (e.g.,
$P=\pi^\pm,\pi^0,\pi^0)$,  
one finds: 
\begin{multline}
\Br(V^\pm\to P A')\simeq \epsilon^2\times \Br(V^\pm\to P \gamma) 
\\ 
\times \frac{(m_V^2-m_{A'}^2-m_P^2)^2\sqrt{(m_V^2-m_{A'}^2+m_P^2)^2-4m_V^2m_P^2}}{(m_V^2-m_{A'}^2)^3}\,,
\label{vector}
\end{multline}
where $m_V$ is the mass of the decaying vector meson and $m_P$ is the mass of
the pseudoscalar meson.

We are interested only in the mesons with a  sufficient decay 
branching ratio. Another 
obvious requirement is that those same mesons should be produced in sufficient
quantities in the SHiP setup. Therefore, in what follows we account only
for $\pi^0$ and $\eta^0$ contributions.

\section{Hidden photon decay pattern}
\label{sec:hidden photon-decays}

\begin{figure}
\centerline{
\includegraphics[scale=0.35]{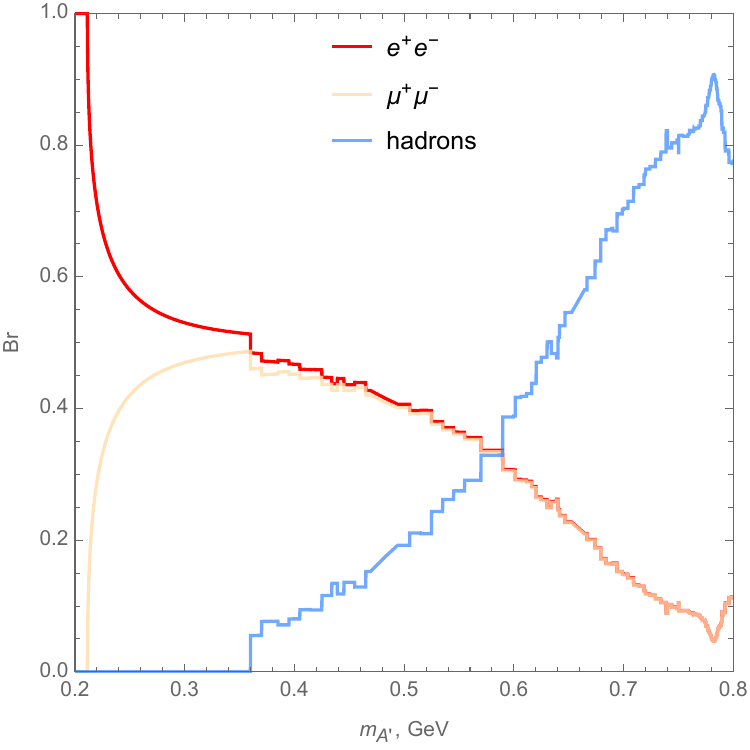}
\includegraphics[scale=0.35]{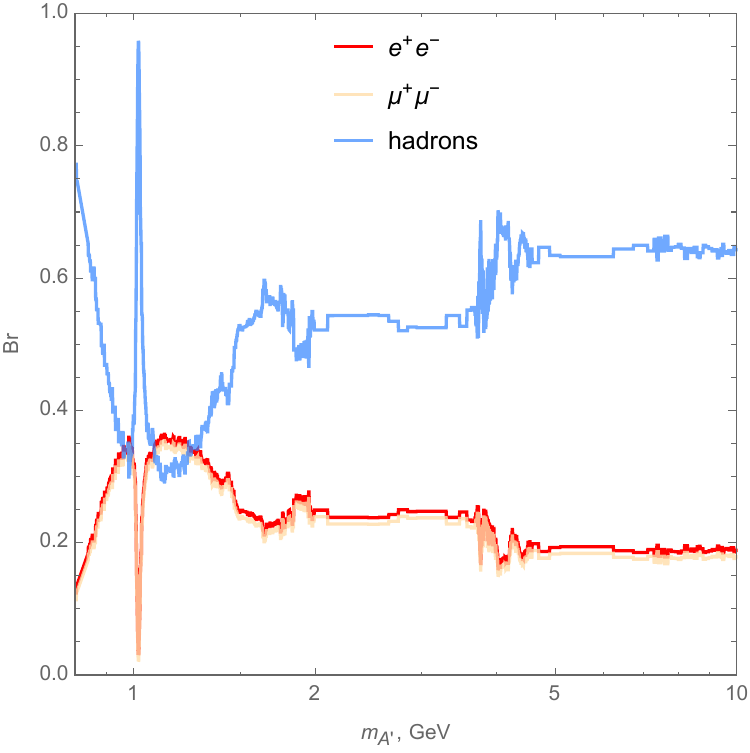}
}

\hskip 0.05\columnwidth 
\includegraphics[width=0.9\columnwidth]{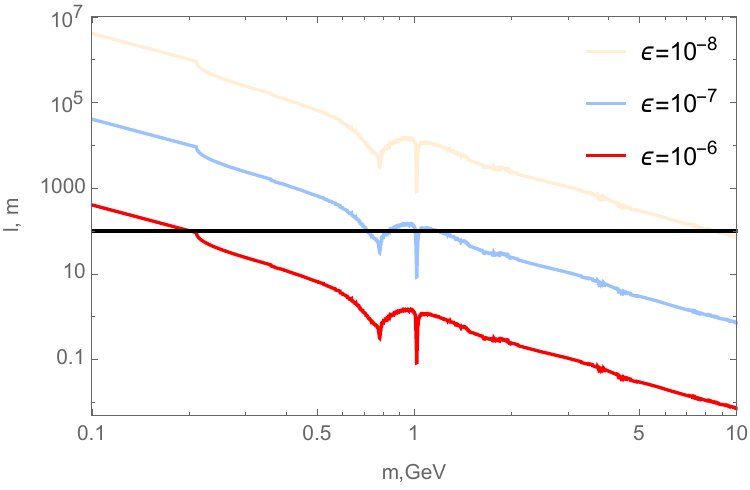}
\caption{{\it Top panels:} 
branching ratios of $A'$ into $e^+e^-$, $\mu^+\mu^-$ and
  hadrons. {\it Bottom panel:} 
decay length of $A'$ with energy 50 GeV; horizontal line represents
  the charachteristic length scale  $100$\,m, particles of much shorter
  decay length do not reach SHiP detector. 
\label{Br_A}}
\end{figure}

The photon-paraphoton mixing $\epsilon$ is responsible for hidden photon
decays into pairs of charged SM particles. 
The partial decay width into a lepton pair is given by \cite{Blumlein:2013cua}
\begin{equation*}
\Gamma_{A'}^{l^+l^-}=\frac13\alpha_{\text{QED}}m_{A'}\epsilon^2\sqrt{1-\frac{4m_l^2}{m_{A'}^2}}\left(1+\frac{2m_l^2}{m_{A'}^2}\right),
\end{equation*}
where $m_l$ is the lepton mass. 
The partial decay width into hadrons can be estimated as 
\begin{equation}
\Gamma_{A'}^{\text{hadrons}}=\frac13\alpha_{\text{QED}}m_{A'}\epsilon^2 \cdot R(m_{A'}),
\end{equation}
where
\begin{equation}
R(\sqrt{s})=\frac{\sigma(e^+e^-\to \text{hadrons})}{\sigma(e^+e^-\to \mu^+\mu^-)}
\label{R}
\end{equation}
is the energy- ($\sqrt{s}$-) dependent ratio \cite{Agashe:2014kda}. 
Resulting branching ratios for the three relevant 
channels are shown in Fig.\ref{Br_A}. 

Neglecting possible invisible decay modes (e.g., those associated
with decay into hidden sector particles if they exist) one 
has the following for the total decay width: 
\begin{equation*}
\Gamma^\text{tot}_{A'}=\Gamma_{A'}^{e^+e^-}+
\Gamma_{A'}^{\mu^+\mu^-}+\Gamma_{A'}^{\text{hadrons}}.
\end{equation*}
Thus the $A'$ decay length reads (see also Fig.\,\ref{Br_A}) 
\begin{equation}
 \gamma c \tau_{A'}=\frac{c\gamma}{\Gamma^\text{tot}_{A'}},
 \end{equation} 
 with the $\gamma$ factor in the laboratory frame, $\gamma=E_{A'}/m_{A'}$.  

\section{Signal event rate}
\label{Sec:Signal} 

\subsection{Hidden photon decays inside the SHiP detector}
\label{sec:hidden photon-signal}

The probability for $A'$ to decay inside 
the fiducial volume of the detector reads 
\begin{multline}
w_{\text{det}}\equiv w_{\text{det}}(E_{A'},m_{A'},\epsilon)=
\exp\l -l_\text{sh}/(\gamma(E_{A'})c\tau_{A'}) \r 
\label{bb}
\\ \times \left [ 1-\exp\l -l_\text{det}/(\gamma(E_{A'})c\tau_{A'}) \r \right ],
\end{multline}
where $l_\text{sh}$ is the muon shielding length (60\,m 
for SHiP \cite{Bonivento:2013jag}) and
$l_\text{det}$ is the length of the detector fiducial volume (50\,m).

The proposed fiducial volume of the SHiP detector is the 50\,m-length
cylindrical vacuum vessel of 5\,m diameter.  To estimate the expected
number of events we use a more conservative volume that is the cone with
the vertex in a target pointing to the 5 m-diameter circle {\it at the
  end of the fiducial volume.}  Thus, we select the hidden photons with momenta
inside that cone, which means 
\begin{equation}
\frac{|p_\perp|}{p_\parallel}<\frac{2.5}{60+50}\equiv\theta_0.
\label{cut}
\end{equation}
We apply this cut to the momenta of hidden photons produced via
bremsstrahlung. The corresponding restriction on momentum space 
refers to  $f_{\text{fiducial}}$ in \eqref{fcut}. 

\begin{figure}[!htb]
\centerline{
\includegraphics[width=0.5\columnwidth]{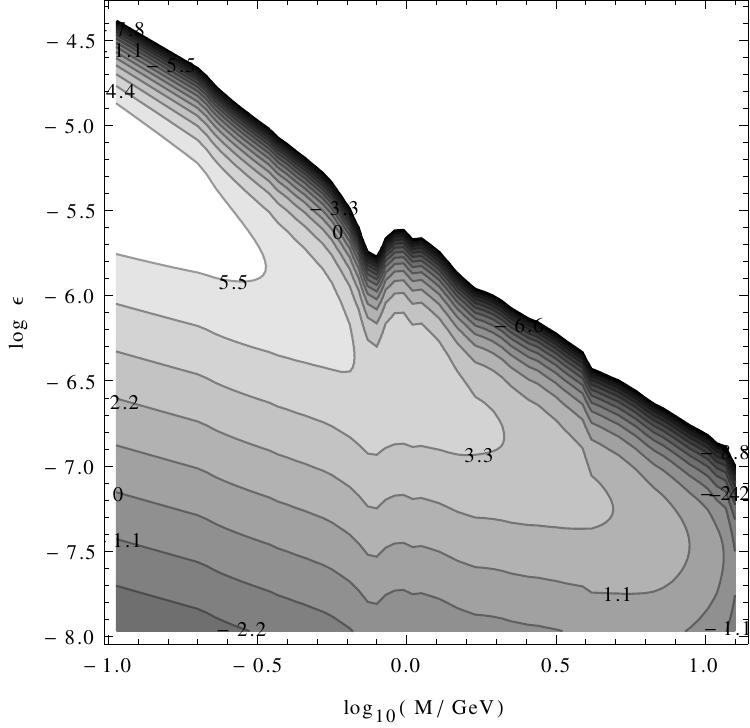}
\includegraphics[width=0.5\columnwidth]{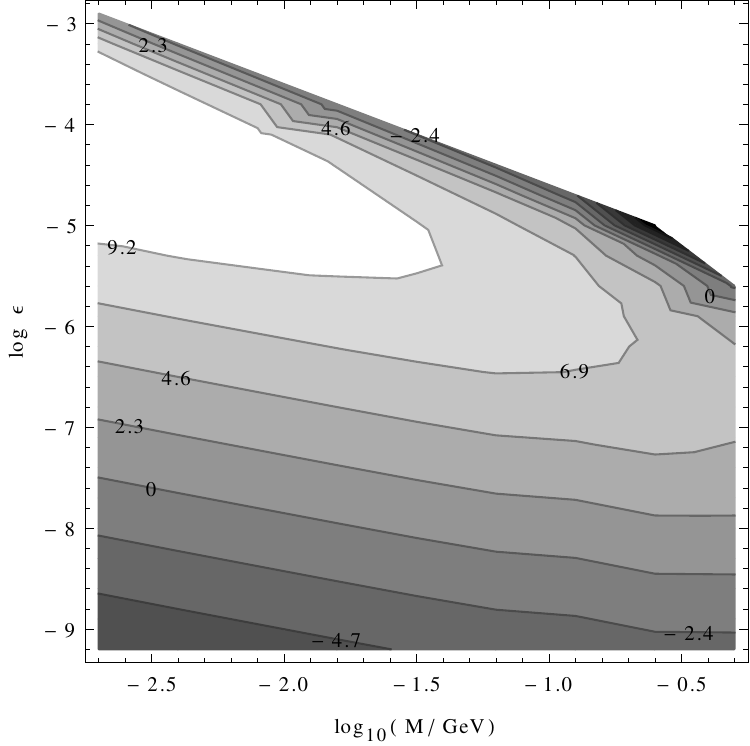}
}
\caption{The $\log_{10}$ of expected number of $A'$ in the detector 
fiducial volume: proton bremsstrahlung 
({\it left panel}) and meson decays
  ({\it right panel}). For the proton contribution we set the Dirac form
factor to unity.}
\label{ProtonSig}
\end{figure}

For the proton channel one integrates the differential flux
\eqref{prod.rate} with $w_\text{det}$ over the region limited by 
\eqref{cut}. The number of $A'$ decays in the detector is then given
by 
\begin{equation}
N_\text{sig}=N_{\text{POT}}\frac{\sigma_{pp}(\!s\!)}{\sigma_{pp}(\!s'\!)}
\!\!\int \!\!\!dz dp_{\perp}^2 w_{ba}(\!z,p^2_\perp\!)
 w_{\text{det}}
f(\!z,p_\perp^2\!),
\end{equation}
where $f(z,p_\perp^2)$ is defined in eq.\,\eqref{fcut}. We assume
that the total number of protons on target will be $N_{\text{POT}} =
10^{20}$. The expected number of events is shown in
Fig.\,\ref{ProtonSig} (left panel). 
The feature at $0.8\,\GeV$ is due to the $\omega$-meson
peak in ratio\,\eqref{R}.

To roughly estimate contribution of secondary protons we assume that
their average energy is $(400\GeV/\text{multiplicity})$ and apply the
same procedure described above.

\subsection{Monte Carlo simulation}
A proton hitting the target initiates a shower of secondary particles
inside the target and muon dump. To precisely account for energy and
angular distributions of secondaries, we simulate both the hadronic and 
the electromagnetic component of the showers. We use the 
Geant4 simulation toolkit 
\cite{geant4} for this purpose. 

Since the SHiP muon dump design is under discussion we choose the most 
simple geometry for the proton target and muon dump:  
rectangular parallelepiped of $40\times40$\,cm$^2$ front 
section and $60$\,m length, 
made of tungsten.  Protons of $400$\,GeV are directed to the target. 

We set Geant4 standard options for particle lists and electromagnetic
processes.  The only essential choice is the selection of the inelastic
hadronic processes generator. We apply 2 recomended
options---FTFP\_BERT (FRITIOF model) and QGSP\_BIC (quark gluon string
model)---and find significant difference in both proton interacton
cross section and the energy-angular dependence of produced secondary
particles. We compare the inclusive cross section with
ref.~\cite{AguilarBenitez:1991yy} and accept the usage of the QGSP\_BIC
model.

\begin{figure}[!htb]
\includegraphics[width=0.5\textwidth]{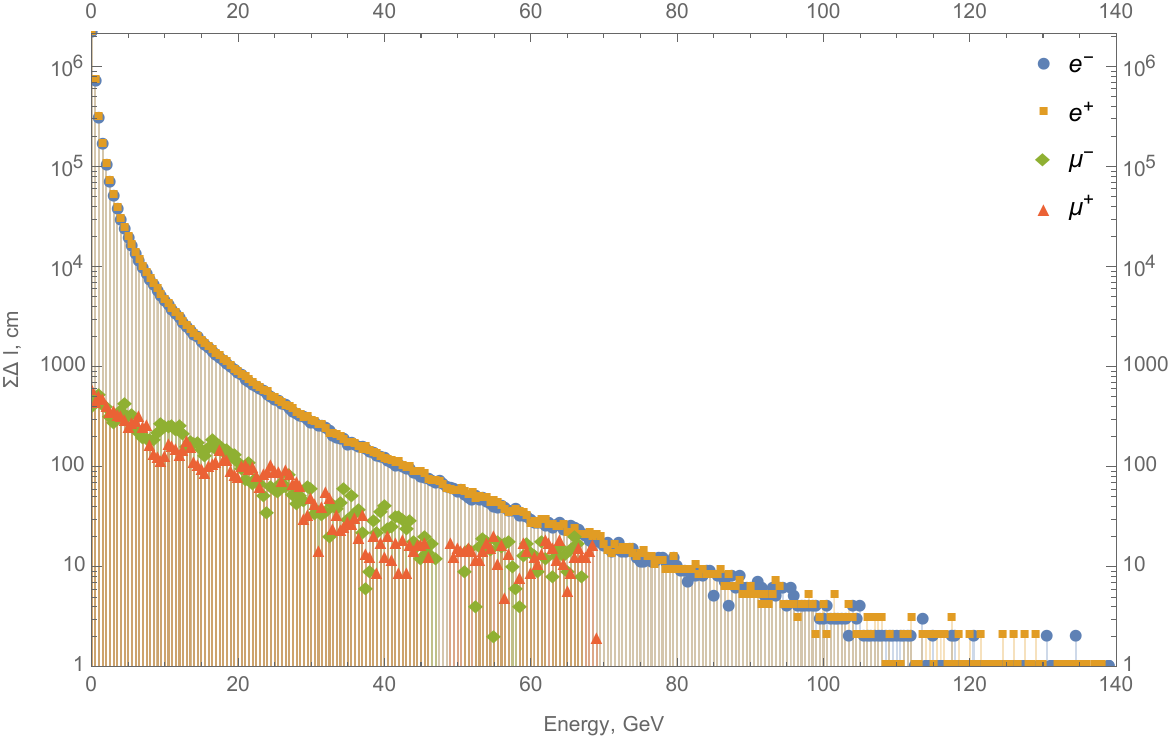}
\caption{The overall length of 
tracks of the particles with energies in a given bin. Result of 160 000 protons on target simulation.}
\label{DeltaLDistri}
\end{figure}

Cases of short-lived and long-lived particles are 
processed differently as follows:
\begin{itemize}
\item 
for each meson produced in the program we save its type ($\pi^0$, $\rho^0$
or $\eta$), and its 3-momentum $p$, which is the energy and angle to beam axis
$\theta_z$;
\item 
for leptons we fill in the histograms with the total lengths of the tracks
of the particles with given energy $E_0$: $L(E_0)=\sum\Delta
l(E\!\!=\!\!E_0,\theta\!\!\leq\!\!\theta_0)$, where the sum runs over all
segments $ \Delta l(E,\theta)$ of all tracks of the particles in the
shower inside the material.  We chose binning $0\dots400$\,GeV with bin
step $0.5$\,GeV.  In this way we get four distinct histograms for
$\mu^+,\ \mu^-,\ e^+,\ e^-$ shown in Fig.\,\ref{DeltaLDistri}.
\end{itemize}

To accelerate simulation we apply cut $p > 0.01$\,GeV to all
particles. To estimate the
dark photon rates, 160K of primary protons have been generated.

\subsection{Analysis}

Using distributions described above, we calculate the number of signal
events inside the detector.  For the meson channel the number of events is
simply $N_{sig} = \sum_{p} N_m(p) \Br(m\to A' X) w_{det} C_{cut},$
where $p$ is the meson momentum, $m$ stands for meson type, e.g. $\pi^0,
\eta^0$, $N_m(p)$ is the amount of mesons with given 3-momentum, $\Br$ is
defined in \eqref{neutral}, \eqref{vector}. The Momentum-dependent
numerical coefficient $C_{cut}$ accounts for the fraction of $A'$
traveling through the fiducial volume.  
In $N_m$ we count only mesons with $\theta_z<\theta_0$.  
The resulting number of signals is presented in Fig.\,\ref{ProtonSig}.

For secondary electrons propagating in the dump medium 
one estimates the number of produced hidden photons as
\begin{equation}
\mathcal{N}\simeq n_a \sum_{E_0} \sigma_{forward}L(E_0),
\end{equation}
where $n_a$ is the nuclei number density, $\sigma_{forward}$ is
the differential cross section \eqref{ecrosssection} integrated with
respect to the geometry constraint, and $E_0$ is the electron 
energy. 

\begin{figure}[!htb]
\includegraphics[width=\columnwidth]{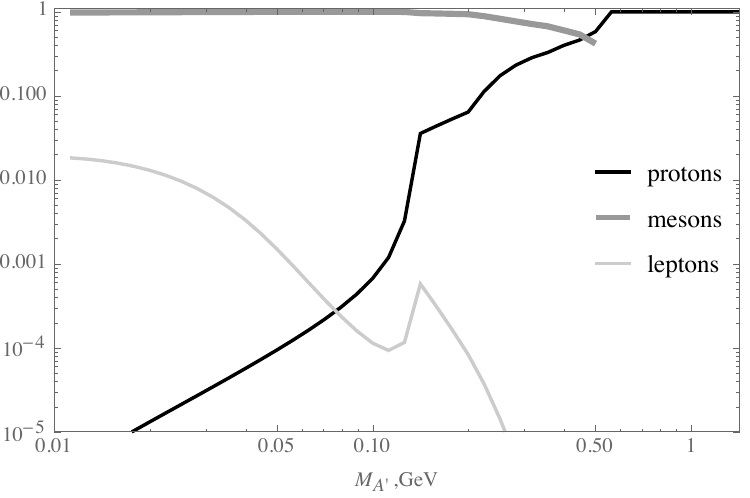}
\caption{The partial contribution of proton, 
meson and electron channels
  to the total number of events.  The kinks on the proton and electron
  lines near the pion mass correspond to the kinematic cut of the pion
  branching ratio. 
\label{CombinedLimits}}
\end{figure}

Using this distribution we calculate the number of $A'$ decays inside the
detector.  Partial contributions of the proton, lepton and meson
channels to the signal event number are shown 
in Fig.\,\ref{CombinedLimits}.  
At small masses meson decays
dominate. The proton bremsstrahlung channel starts to dominate when
$m_{A'}>m_{\eta^0}$ since heavier $A'$ could not be produced in meson
decays in sufficient amounts.  In the whole mass region 
the lepton bremsstruhlung contribution is neglidibly small and, hence, omitted.

To define the domain of parameters $m_{A'}$ and $\epsilon$ where SHiP will
be sensitive to hidden photons we assume $10^{20}$ POT, neglect
background, and adopt the Poisson statistics which tell us that no events
while three events are expected implies an exclusion at 95\% confidence
level. The exclusion limits are shown in Fig.\,\ref{limits}. The dotted
black line outlines proton contributions of initial and secondary
protons, while solid black line refers to the conservative constraint
accounting for the Dirac form factor of proton, see
Sec.\,\ref{subsec:proton}.  We argue that the actual constraint is in
a region between the dotted and solid lines.  As shown in
Sec.\,\ref{sec:hidden photon-decays}, the dark photon lifetime is
proportional to $\epsilon^{-2}$, thus, the upper border of the region in
Fig. \ref{limits} corresponds to a quick decay of the dark photon.
For the values of $\epsilon$ above the upper line, the hidden photons
decay within the shielding.  The lower border of the region
corresponds reciprocally to a slow decay.

\section{Discussion}
\label{sec:Conclusion}

To summarize, we have estimated the hidden photon signal rate expected
in the SHiP experiment, and outlined the region in model parameter
space (see Fig.\,\ref{limits}), where three paraphoton decays in a 50\,m-length
detector are expected for each $10^{20}$ protons on target. The
results are rather {\it conservative} and may be improved by taking
into account: {\it (i)} $\rho$, $\omega$ and other short-lived hadron
contributions to hidden photon production via decay, {\it (ii)}
$\pi^\pm$ and other long-lived hadron contributions to hidden particle
production via bremsstrahlung, {\it (iii)} quark bremsstrahlung
contribution. However, since the number of signal events scales with mixing
as $\epsilon^4$, we expect no significant change in our
results. As the next steps, the real experiment geometry, 
detection efficiency, and background events
must be analyzed to find the sensitivity of SHiP to the hidden photon
models. 

\vspace{0.2cm}
We thank W.\,Bonivento, A.\,Golutvin, R.\,Lee, N.\,Serra 
and M.\,Shaposhnikov for useful discussions. 
This work has 
been supported by Russian Science Foundation grant 14-22-00161. 


\bibliography{refs}

\end{document}